\documentclass[prb,twocolumn,,amsmath,amssymb,floatfix]{revtex4}
\usepackage{graphicx}
\usepackage{bm}

\usepackage{amssymb}
\usepackage{color}

\newcommand{\be}{\begin{equation}}
\newcommand{\ee}{\end{equation}}

\def \be{\begin{equation}}
\def \ee{\end{equation}}
\def \ba{\begin{array}}
\def \ea{\end{array}}
\def \bea{\begin{eqnarray}}
\def \eea{\end{eqnarray}}
\def \nn{\nonumber}

\def \etal{{\it {et al}}}

\def \a{{\alpha}}
\def \t{{\theta}}
\def \b{{\beta}}

\def \D{{\Delta}}
\def \d{{\delta}}
\def \w{{\omega}}

\def \nd{{^{\vphantom{\dagger}}}}
\def \yd{^\dagger}
\def \av#1{{\langle#1\rangle}}
\def \ket#1{{\,|\,#1\,\rangle\,}}

\begin{document}

\title{Hidden order in one dimensional Bose insulators}

\author{Emanuele G. Dalla Torre$^1$, Erez Berg$^{1,2}$, Ehud Altman$^1$}
\affiliation {$^1$ Department of Condensed Matter Physics,
Weizmann Institute
of Science, Rehovot, 76100, Israel\\
$^2$ Department of Physics, Stanford University, Stanford, CA
94305-4045, USA}

\begin{abstract}
We investigate the phase diagram of spinless bosons with long range
($\propto 1/r^3$) repulsive interactions, relevant to ultracold
polarized atoms or molecules, using DMRG. Between the two
conventional insulating phases, the Mott and density wave phases, we
find a new phase possessing hidden order revealed by non local
string correlations analogous to those characterizing the Haldane
gapped phase of integer spin chains. We develop a mean field theory
that describes the low energy excitations in all three insulating
phases. This is used to calculate the absorption spectrum due to
oscillatory lattice modulation. We predict a sharp resonance in the
spectrum due to a collective excitation of the new phase that would
provide clear evidence for the existence of this phase.
\end{abstract}

\date{\today}

\maketitle

{\em Introduction --} Systems of ultracold atoms in optical lattices
hold significant promise in the study of correlated quantum matter.
Experimental realization of the Bose-Hubbard model has been an
important step in this direction, not least because it facilitated
the first observation of a quantum insulating state of
bosons\cite{greiner}. Localization of bosons in the Mott insulator
is driven by strong on site interactions leading to an extremely
simple state, well described by a site-factorizable mean field
wave-function. An interesting question is what other phases, perhaps
with non trivial structure, may be stabilized by longer range
interactions\cite{goral,pai,barnett,senthil,buchler}. This issue has
been given added urgency by recent advances in trapping and cooling
of atoms\cite{dipolar-atoms} and molecules\cite{dipolar-molecules}
with large dipole moments.

Ultracold dipolar bosons, in one dimensional optical lattices can be
described by the hamiltonian

\be H=-t\sum_{i}(b\yd_i b\nd_{i+1}+H.{\rm c}.)+{U\over 2}\sum_i
n_i(n_i-1)+\sum_{i,r>0}{V\over r^3}n_i n_{i+r}, \label{H} \ee where
we assumed that the dipoles are polarized by an external field
perpendicular to the lattice. The three parameters appearing in the
hamiltonian can be tuned independently in experiments\cite{goral}.
The conventional phases of this system at {\em integer filling} are
known from mean field studies\cite{goral,pai}. They include the
Mott insulator (MI) at large $U$, a density wave (DW) for large $V$,
and a superfluid (SF) for large $t$.

In this letter we use the Density Matrix Renormalization Group
(DMRG) method \cite{white} to show that a new gapped insulating
phase with non trivial structure obtains in a wide parameter regime
between the two conventional insulators. The new phase is separated
from the conventional insulators by lines of second order
transitions and is found to possess particularly subtle order that
is revealed only by highly non local string correlation functions.
Correlations of similar nature exist in the Haldane
gapped\cite{haldane} phase of quantum spin-1
chains\cite{rommelse,KennedyTasaki}. We shall therefore term the new
phase as the Haldane Bose insulator (HI). The analogy can be made
more explicit by truncating the Hilbert space of the Bose system to
three occupation states per site (for example, for a system with
${\bar n}=1$ particles per site, we keep only the occupation states
$n=0,1,2$ for every site). This defines an effective spin-1 model
with $S^z_i=n_i-{\bar n}$ and $\bar n$ the average filling. In this
space the DW phase corresponds to antiferromagnetic ordering of the
pseudospins in the $z$ direction. The MI ground state, on the other
hand, includes a large amplitude of the state with $S^z_i=0$ on
every site and small admixture of states containing tightly bound
particle-hole fluctuations ($S^z=\pm 1$ on nearby sites). The
Haldane phase may also contain mostly sites with $S^z=0$ at large
$U$, but the ordering of the fluctuations is unusual. The string
correlations imply that particle and hole fluctuations appear in an
alternating order along the chain separated by strings of zeros of
arbitrary length\cite{rommelse,KennedyTasaki}. The truncation to
three states is justified only at large $U$ when fluctuations in
site occupancy are strongly suppressed. However, from the numerical
results we conclude that the long ranged string order of the actual
bosons survives fluctuations in the occupancy beyond the effective
spin-1 description.

After a description of the DMRG results, we will show how the new
phase can be detected experimentally. The challenge lies in the fact
that standard experimental probes couple to local observables and
are blind to the highly non local string correlations. We shall
therefore look for signatures of the new phase in the excitation
spectrum rather than the ground state properties by considering the
response to lattice modulation\cite{Esslinger}. We develop an
approximation scheme that enables us to calculate the response
functions in all three insulating phases. We predict a sharp
resonance in the absorption due to a neutral collective mode
($S^z=0$ in the pseudospin terminology), that is special to the
Haldane insulator and can serve to identify this phase.

{\em Numerical Results --} We investigate ground state and lowest
excitations of the hamiltonian (\ref{H}) in a space of the
parameters $U$ and $V$ using the DMRG algorithm\cite{white}.
Throughout we consider a filling of $\bar n=1$ particles per site in
the ground state. The maximal length of the system is $N=256$ sites,
and up to $M=200$ states are kept per block. In most of the
parameter regime we found that the results do not change
significantly when the cutoff in boson occupation number per site is
increased beyond 4. The $1/r^3$ interactions was included up to the
next nearest neighbor range. Open boundary conditions were used, and
opposite external chemical potentials were applied on the first and
last site in order to lift the ground state degeneracy in the HI and
DW phases.

\begin{figure}[h]
\centering \includegraphics[scale=0.5]{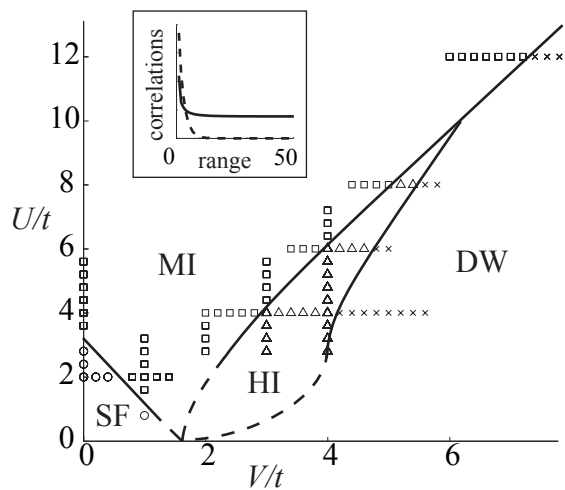}
\caption{\label{phase} Phase diagram of the hamiltonian (\ref{H}) in
the $(U,V)$ plane, obtained from DMRG. The phases that appear in the
diagram are: Superfluid (SF,o), Mott insulator (MI,$\square$),
density wave (DW,x) and Haldane insulator (HI,$\triangle$). INSET:
density wave (dashed line) and string (solid line) correlations at a
particular $(U,V)$ point (see (\ref{DWCorr},\ref{StringCorr})). This
point is in the HI phase, as can be seen from the fact that string
correlations do not decay, whereas DW correlations decay rapidly.}
\end{figure}

\begin{figure}[h]
\centering \includegraphics[scale=0.5]{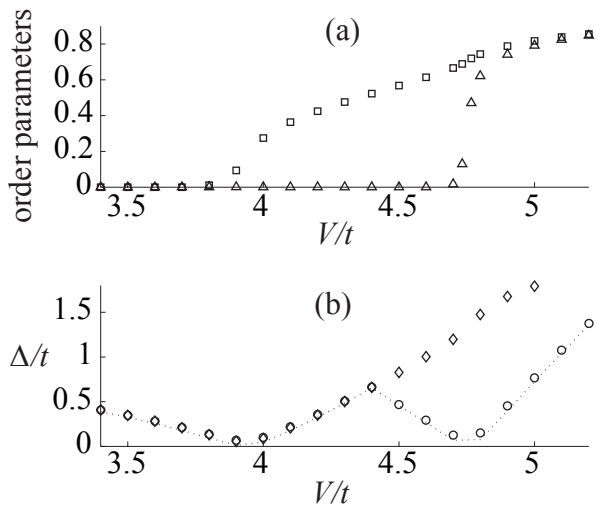} \caption{ (a): DW
order parameter ($\triangle$) and string order parameter ($\square$)
as a function of $V$ along the line $U=6t$. (b): Gaps to neutral and
charged excitations, $\Delta_0$ (o) and $\Delta_1$ ($\diamond$)
respectively (see text) along the same line.} \label{gap}
\end{figure}

The phase diagram of (\ref{H}) in the $(U,V)$ plane is shown in Fig.
\ref{phase}. The nature of the phases in the DMRG simulation was
elucidated by a direct calculation of the ground state correlation
functions:
\begin{eqnarray}
R_{\text{SF}}\left(\left\vert
i-j\right\vert \right) &=&\left\langle
b_{i}^{\dagger }b_{j}\right\rangle  \label{SFCorr} \\
R_{\text{DW}}\left( \left\vert i-j\right\vert \right) &=&\left(
-1\right) ^{\left\vert i-j\right\vert }\left\langle \delta
n_{i}\delta
n_{j}\right\rangle  \label{DWCorr} \\
R_{\text{string}}\left( \left\vert i-j\right\vert \right)
&=&\left\langle \delta n_{i}e^{i\pi \sum_{k=i}^{j}\delta
n_{k}}\delta n_{j}\right\rangle \label{StringCorr}
\end{eqnarray}
Here $\delta n_i \equiv n_i-{\bar n}$. The superfluid phase is
characterized by a power law decay of $R_{\text{SF}}\propto
|i-j|^{-1/2K}$ with Luttinger parameter $K\ge 2$. In the MI phase, all the correlation
functions decay exponentially to zero. In the DW phase $R_{DW}\left(
\left\vert i-j\right\vert \right)$ approaches a constant at long
distances (note that also $R_{\text{string}}\left( \left\vert
i-j\right\vert \right) \rightarrow const.\neq 0$), and the lattice
translation symmetry is spontaneously broken. In analogy to the
$S=1$ XXZ chain \cite{rommelse,Schulz,KennedyTasaki,numerics_xxz},
we expect the appearance of another phase (the Haldane phase)
between the MI and DW phases. This phase is characterized by $R_{\text{%
string}}\left( \left\vert i-j\right\vert \right) \rightarrow
const.\neq 0$, while the DW correlations decay exponentially. Unlike
the density wave phase, this phase does not break the lattice
translation symmetry. It does, however, break a hidden $Z_{2}$
symmetry related to the string order parameter
\cite{tasaki,KennedyTasaki}. Fig. \ref{gap}(a) presents an example
of how the phase boundaries were determined: we show the string and
DW order parameters (defined as the square root of the asymptotic
values of the corresponding correlation functions) as a function of
$V$ along the line $U=6t$. The phase transitions from MI to HI and
from HI to DW are clearly visible and seem to be of second order.

The phase diagram does not change qualitatively if we keep only
the nearest neighbor interactions in (\ref{H}). However, further
range interactions act to frustrate the DW order and thereby widen
the domain of the HI phase. We note that previous DMRG studies of 
the Bose-Hubbard model with nearest neighbor interaction [17] did 
not look for the string correlations and therefore did not find the subtle HI phase.

In addition to the ground state, the energies of the first few
excited states were calculated. The gap to the first excited state
with the same number of particles as the ground state,
$\Delta_{0}=E^{(1)}_{\delta n=0}-E^{(0)}_{\delta n=0}$, was
calculated by targeting also the first excited state in the DMRG
calculation. Here $\delta n$ is the number of particles relative
to a state with exactly ${\bar n=1}$ particles per site. The
charge gap of the system $\Delta_{1}=E^{(0)}_{\delta
n=1}+E^{(0)}_{\delta n=-1}-2E^{(0)}_{\delta n=0}$ was calculated
by targeting the ground states of the $\delta n=\pm 1$ sectors.

The gaps $\Delta_0, \Delta_1$ along the line $U=6t$ in the $(U,V)$
plane are shown in Fig. \ref{gap}(b). At the transition point
between the MI and HI phases, both $\Delta_0$ and $\Delta_1$ vanish,
while at the transition between the HI and DW phases only $\Delta_0$
vanishes. This indicates that both transitions are second order, but
the nature of the critical excitations at the transition is
different.

In the MI phase, the lowest excitations are particle and hole
excitations. This is clear from the fact that $\Delta_1=\Delta_0$,
i.e., the first excitation with $\delta n=0$ is an unbound
particle-hole pair. The gaps to these particle and hole excitations
vanish at the transition to the HI. The Haldane phase displays
another low-energy excitation. In a certain region of the phase
diagram $\D_0$ goes below the charge gap indicating the presence of
a genuine {\em neutral} mode with $\delta n=0$. In Fig. \ref{gap}(b)
the crossing point
between the two states is clearly seen as a cusp in the $\Delta_0$
curve near the middle of the HI phase region. At the transition to
the DW phase, only the neutral gap vanishes. The presence of the
neutral mode seems to be characteristic of the Haldane phase, and
can be used to detect it, as will be discussed below.

{\em Experimental detection --} In the numerical simulation we were
able to establish the presence of the string ordered insulating
phase HI by measuring the non local ground state correlations
(\ref{StringCorr}) directly. By contrast, experimental probes
naturally couple to local operators, such as the charge density.
Our detection strategy will focus on probes of the excitation
spectrum, which may exhibit distinct (albeit indirect) signatures of
the HI phase.

Let us consider in some detail the response to
parametric excitation of the optical lattice. This technique was
used successfully to probe the excitations in both the MI and SF
phases\cite{Esslinger}. The idea is to apply a periodic modulation
of the lattice intensity. Within the lowest Bloch band this
perturbation simply corresponds to uniform modulation of the
hopping matrix element in (\ref{H}). The perturbed hamiltonian can
be written as $H+h \cos(\w t)\hat T$, where $\hat T$ is the
kinetic energy operator ${\hat T}=\sum_i b\yd_i b\nd_{i+1}+H.{\rm
c}.$. Within linear response, the absorption rate is
\be I(\w)\sim
\sum_\alpha \left|\langle \psi_\alpha | \hat{T} | \psi_0 \rangle
\right|^2\delta(\w_{\alpha0} - \w) \label{Iw}
\ee
We shall
calculate $I(\w)$ within the effective spin-1 hamiltonian
\be
H=J\sum_{i}\left(S^+_iS^-_{i+1}+{\mbox H.{\rm c}.}\right)+\sum_i
VS^z_i S^z_{i+1} +{U\over 2}(S^z_i)^2, \label{spin1}
\ee
using the
corresponding perturbation operator ${\hat T}=\sum_i
S^+_iS^-_{i+1}+{\mbox H.{\rm c}.}$. The mapping to a spin model
amounts to projecting (\ref{H}) on the subspace including only
three occupation states per site ($\d n_i=0,\pm 1$). This is
justified in the insulating phases at large $U$, where multiple
particle or hole occupancy is suppressed. The
$U$ term in
(\ref{spin1}) corresponds to on site interaction, $V$ to nearest
neighbor interactions, and $J$ to boson hopping. The
projection of the hoppnig term in (\ref{H}) gives another
contribution to the spin model, that breaks the particle-hole
symmetry. Omitting this term, as we have done in (\ref{spin1}),
does not lead to qualitative changes in the phase diagram.
The advantage in using the model
(\ref{spin1}) is that it is amenable to a mean field theory,
developed by Kennedy and Tasaki\cite{KennedyTasaki}, that captures
all three insulating phases.

Kennedy and Tasaki introduced a non local unitary operator that
transforms the string correlation (\ref{StringCorr}) to
conventional spin correlations $\av{S^z_iS^z_j}$, which admit a
local mean field treatment. At the same time the hamiltonian
(\ref{spin1}) assumes a rather unusual, but nevertheless local form:
\bea \tilde{H} &=&-J\sum_{j} S^x_{j} S^x_{j+1} - S^y_{j} \exp(i \pi
S^z_j + i \pi S^x_{j+1})
S^y_{j+1} \nn\\
&&- V\sum_{j} {S}^{z}_j {S}^{z}_{j+1} + {U\over 2} \sum_j (
{S}^{z}_j)^2 \label{Htilde} \eea
The $U(1)$ symmetry of the original
hamiltonian seems to have been broken down to $Z_2\times Z_2$. In
fact the full $U(1)$ symmetry is preserved but is generated by a
highly non-local operator in the new representation.

\begin{figure}[t]
\centering \includegraphics[scale=0.5]{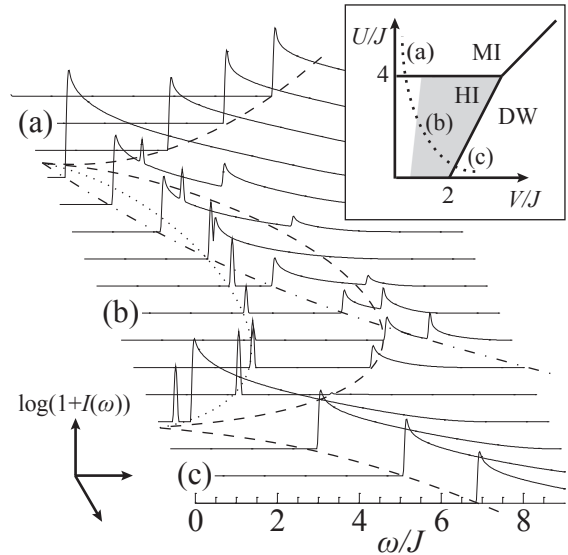}
\caption{Absorption spectrum
(\ref{Iw}) calculated from the mean field theory along the dotted line in the
inset (mean field phase diagram of
(\ref{Htilde})). The dotted line in the spectra marks the position of a
quasi-particle peak seen only in the HI phase (b). The
dashed and dash-dotted lines mark the edges of two particle
continua.} \label{fig:Iw}
\end{figure}

The mean field theory of (\ref{Htilde}) consists of finding the best
site-factorizable wave-function for this hamiltonian. The solutions
are of the general form\cite{KennedyTasaki} $~\ket{\Psi}=\prod_i
(\cos\t\ket{0}_i\pm\sin\t\ket{\pm 1}_i)$. The resulting phase diagram
is shown in the inset of Fig. \ref{fig:Iw}.
In the Haldane phase
$0<\t<\pi/2$ and the ground state is four fold degenerate, which
reflects broken $Z_2\times Z_2$ symmetry. $\t$ is given explicitly
by the  expression $\cos^2\t=(U-2V+4J)/(8J-2V)$. The DW  phase is
characterized by $\sin\t=1$, i.e. doubly degenerate ground state.
Finally in the MI there is a non degenerate ground state with
$\cos\t=1$. The fact that the superfluid phase is pushed to negative
$V$ is an artifact of the projection to three occupation states.

The collective excitations are found, following a standard
scheme\cite{AltmanAuerbach}, by quantizing the small fluctuations
around the mean field minima.
We obtain an effective hamiltonian of the form
$H_{eff}=\sum_{\a k} \w_{\a k}\b\yd_{\a k}\b\nd_{\a k}$ for the two
collective modes $\b_{\a k}$ ($\a=1,2$). Likewise we expand the
perturbation $\hat T$ in terms of the collective mode operators. Finally we
compute $I(\w)$ using (\ref{Iw}).

Fig. \ref{fig:Iw} displays the results at different points along a
line in the $(U,V)$ plane that cuts through the three phases. As
expected we see the gap in the spectrum closing at the
two quantum phase transitions. The most distinct
signature of the Haldane phase is a delta-function peak in the
absorption spectrum corresponding to creation of a single
quasi-particle. This resonance can serve as a "smoking gun"
experimental evidence of the HI phase.
The fact that this quasi-particle couples to
$\hat T$ which does not change particle number
suggests that it must be a neutral excitation (quantum number $\d
n=0$). Similarly we infer that the other quasi-particle, which does
not couple to $\hat T$, is a charged mode. The other two
peaks seen in the spectra are the bottom edges of the two
particle continua corresponding to a pair of "charged" and
a pair of neutral excitations respectively.

While the mean field treatment is approximate and cannot be expected
to capture details such as the location of phase boundaries,
it agrees with the DMRG results on the qualitative features.
In both calculations the signature neutral mode
is below the particle hole continuum in a wide region
inside the HI phase (see Fig. \ref{gap}(b) and Fig. \ref{fig:Iw}, gray region in the inset).
The mean field theory also captures
the fact that only the neutral mode
becomes critical at the transition from the HI to the DW state.
Because of the lattice symmetry breaking in the DW phase,
we expect a neutral excitation with momentum
$k=\pi$ to drop to zero energy at the transition to this phase.
This mode is adiabatically connected to the $S^z=0$ magnon of
the spin-1 Heisenberg chain. However it cannot be identified with the
delta function peak in
Fig. \ref{fig:Iw},
because the operator $\hat T$ couples only to
$k=0$ excitations. The delta function peak is therefore
interpreted as a $k=0$ bound state of two such $k=\pi$ excitations.
A single $k=\pi$ excitation will show up as a sharp peak
in Bragg spectroscopy\cite{Stenger}, in which the lattice is modulated with another laser
of wave-vector $k=\pi$.


We now discuss the possibility of realizing the HI phase
experimentally in a system of ultra-cold atoms in a one
dimensional optical lattice. From Fig. \ref{phase} we see that this
requires $V\sim U$. This condition can be expressed in terms of the
elastic scattering length, $a_s$, and the effective scattering
length associated with a dipole $d$, $a_d=- m d^2 / 4\pi \hbar^2$.
For a typical optical lattice potential, with wavelength $k_L$ and effective
depth of the order of $10E_R$ ($E_R = (\hbar
k_L)^2 / 2m$), we estimate\cite{jaksch} $U \approx 12 (k_L a_s) E_R$.
The nearest neighbor interaction energy is $V = d^2/ r^3 \approx d^2
k_L^3 / \pi^3$ and therefore the condition $V \sim U$ is equivalent
to $|a_d| \sim 30 a_s$. Since the typical scattering length of atoms
is of the order of tens of angstroms, we need an effective dipole
scattering length of order $1000$ \AA.
Dipoles of this magnitude can be obtained with ultracold
dipolar molecules \cite{baranov}.
Alternatively it might be possible to use $Cr$ atoms with $a_d\approx 6 \AA$ if
the other energy scales $t$ and $U$ are reduced proportionally. Of course the temperature
would have to be lowered to the same scale, below the gap $\D\sim t\sim V$ seen in Fig. \ref{gap}(a).
This is achieved automatically by adiabatic
cooling as the optical lattice potential is increased slowly to the desired value.
The fact that the HI phase is incompressible implies that in the
presence of a harmonic confinement it will form a plateau structure,
similar to the usual Mott insulator.

{\em Discussion and conclusions --} We have shown that addition of
further range interactions to the Bose-Hubbard model can give rise
to a new insulating phase with non local string order, analogous to
the Haldane gapped phase of integer spin chains. The new phase can
be realized in systems of ultracold dipolar atoms or molecules in
optical lattices. We predicted unique signatures
of the new phase in the parametric excitation spectra.
These include low energy critical excitations
that appear near the phase transitions to the conventional phases,
and a sharp resonance in the response seen only in the new phase.

{\em Acknowledgements --} Useful discussions with D. Arovas, M. Brociner, T.
Esslinger and W. Ketterle are gratefully acknowledged.

\end{document}